# A statistical model approximation for perovskite solid-solutions: a Raman study of lead-zirconate-titanate single crystal

J. Frantti[1], Y. Fujioka[1], A. Puretzky[2], Y. Xie[3], Z.-G. Ye[3] and A. M. Glazer[4]

[1] Department of Applied Physics, Aalto University, FI 00076 Aalto, Finland

[2] Center for Nanophase Materials Sciences, Oak Ridge National Laboratory, Oak Ridge, Tennessee 37831, USA

[3] Department of Chemistry and 4DLABS, Simon Fraser University, Burnaby, BC, V5A 1S6, Canada

[4] Clarendon Laboratory, Department of Physics, University of Oxford, Parks Road, Oxford OX1 3PU, United Kingdom

## Abstract

Lead titanate ($PbTiO_3$) is a classical example of a ferroelectric perovskite oxide illustrating a displacive phase transition accompanied by a softening of a symmetry-breaking mode. The underlying assumption justifying the soft-mode theory is that the crystal is macroscopically sufficiently uniform so that a meaningful free energy function can be formed. In contrast to $PbTiO_3$, experimental studies show that the phase transition behaviour of lead-zirconate-titanate solid solution (PZT) is far more subtle. Most of the studies on the PZT system have been dedicated to ceramic or powder samples, in which case an unambiguous soft-mode study is not possible, as modes with different symmetries appear together. Our Raman scattering study on titanium-rich PZT single crystal shows that the phase transitions in PZT cannot be described by a simple soft-mode theory. In strong contrast to $PbTiO_3$, splitting of transverse *E*-symmetry modes reveals that there are different locally-ordered regions. The role of crystal defects, random distribution of Ti and Zr at the *B*-cation site and Pb ions shifted away from their ideal positions, dictates the phase transition mechanism. A statistical model explaining the observed peak splitting and phase transformation to a complex state with spatially varying local order in the vicinity of the morphotropic phase boundary is given.

## Introduction

Understanding the microscopic mechanism resulting in a phase transition is a central challenge in materials science, as exemplified by ferroelectrics, a classical set of materials characterized by the appearance of a spontaneous polarization (an order parameter) below the transition point separating paraelectric (high-symmetry) and ferroelectric (low-symmetry) phases. The polarization direction can be switched between two equivalent orientation states [1], and it is essentially this property that ferroelectric applications utilize. The nature of the phase transition dictates the number of orientation states available below the transition point, which is one of the prime reasons why the details of the ferroelectric transition are studied. The free-energy is expanded as a function of the order parameter, which allows one to write down an equation of motion and to show that at a continuous phase transition the soft-mode phonon frequency is zero [2]. In the well-known perovskite oxides, $BaTiO_3$ and $PbTiO_3$, it is the transverse optical (TO) Brillouin zone centre mode that softens and results in the formation of a large tetragonal distortion through a first-order phase transition. The soft-mode is Raman-active in the low-symmetry and low-temperature phase, allowing an experimental probing through light scattering techniques.

Lattice dynamical considerations of $BaTiO_3$ and $PbTiO_3$, based on the rigid-ion model [3], is given in ref. 4. An early self-consistent phonon computation conducted for perovskite oxides was able to provide a model for phase transitions from the cubic phase to tetragonal, orthorhombic and rhombohedral phases [5]. An essential feature of this model is that it takes the coupling between strain (exemplified by acoustic modes) and soft-mode into account, which is crucial in tetragonal $PbTiO_3$ as the large $c/a$ axis ratio reveals. Density functional theory revealed that the low-frequency phonon-density-state is large, making the aforementioned coupling plausible [6]. The model also provided an explanation of how the order of the phase transition depends on the coupling. Example systems included $PbTiO_3$, $BaTiO_3$, $KNbO_3$, lead-zirconate-titanate [$Pb(Zr_xTi_{1-x})O_3$, PZT] and $KTa_xNb_{1-x}O_3$ [5]. Anharmonic interactions (coupling) were assumed to be due to short-range forces only. In $PbTiO_3$, the soft mode in the high-symmetry phase is threefold degenerate $T_{1u}$(TO) (there being three equivalent, orthogonal directions) and splits into twofold $E$(TO) and nondegenerate $A_1$(TO) modes in the low-symmetry phase. The model correctly describes the fact that the squared frequency of the $A_1$(TO) mode is proportional to the squared order parameter (or, in other terms, tetragonal strain $c/a$-1), whereas the squared $E$(TO) frequency depends not only on the squared value of the order parameter but also involves strain terms and correlation functions of the acoustic and soft modes. A Raman scattering study [7] on $PbTiO_3$ single crystal was found to be consistent with the model, the softening of the $A_1$(1TO) mode being proportional to $\sqrt{c/a-1}$, whereas the $E$(1TO) mode had a flatter temperature dependence. Though the $A_1$(1TO) peak was actually measured in ref. [7], it was not correctly identified. Owing to the large anharmonic contribution the $A_1$(1TO) mode lineshape is anomalous. Thus, the $A_1$(1TO) mode frequency $\omega(A_1(1\mathrm{TO}))$ was estimated from Merten's equation [8] in ref. 7. It was twenty years later that the $A_1$(1TO) mode was correctly identified and the lineshape was shown to be a superposition of several subpeaks [9]. As a matter of curiosity, the $\sqrt{c/a-1} \propto \omega(A_1(1\mathrm{TO}))$ relation is much improved if the measured $\omega(A_1(1\mathrm{TO}))$ values are used. The coupling between optical and acoustic modes is not limited to soft modes, but strain can couple to other optical Brillouin zone modes [10]. The coupling changes the elastic constants and thus can cause lattice instability, as discussed in ref. 10. The coupling between soft-optical-mode coordinates and strain was implemented into a model Hamiltonian to understand the cell-doubling transitions in $SrTiO_3$ and $LaAlO_3$ in which case the soft mode is a Brillouin zone boundary mode resulting in tetragonal and rhombohedral distortions, respectively [11]. By replacing a fraction of the titanium by zirconium in $PbTiO_3$ the PZT solid-solution is formed [12]. Recently, coupling between the lowest-frequency $A_1$ mode and the first overtone of the oxygen octahedral tilting mode (an antiferrodistortive mode) was reported to yield a splitting of the $A_1$ mode into two in $Pb(Zr_{0.53}Ti_{0.47})O_3$ [13]. Condensation at the Brillouin zone $A$ point, $q = \frac{\pi}{a}\frac{\pi}{a}\frac{\pi}{c}$, of a $B_1$-symmetry mode corresponds to the transition $P4mm \rightarrow I4cm$, which was predicted to occur under 10 GPa hydrostatic pressure in $PbTiO_3$, followed by a condensation of the $E$-symmetry mode further lowering the symmetry to rhombohedral $R3c$ [14].

PZT ($PbZr_xTi_{1-x}O_3$)is among the most intensively studied ferroelectric oxides exhibiting exceptionally high piezoelectric properties when the amount of titanium and zirconium is roughly equal. Essentially because of local-scale disorder numerous space group assignments can be found in the literature. However, the average crystal structures as a function of $x$ can be summarized thus: at room temperature the crystal structure remains tetragonal up to $x \approx 0.52$ at which composition (termed the morphotropic phase boundary, MPB) a rather complex set of phases emerges, including a monoclinic *Cm* phase [15,16,17] and low and high-temperature rhombohedral phases. It is evident that the *Cm* phase exists as judged by several high-resolution powder diffraction studies, but the phase transition mechanism resulting in an average monoclinic phase, and even the stability of the phase, is still under investigation as it is often linked to the extraordinary piezoelectric properties of PZT at the MPB composition. A thermodynamical study showed that an 8th-order Devonshire theory is required to explain the monoclinic phase [18], suggesting that the transition is quite unusual. This was also discussed in ref. 19, where a two-order-parameter thermodynamic model was developed for PZT to account for its peculiar features. At higher values of *x* the structure can, to a good approximation, be described as rhombohedral, there being two variants, with (*R*3*c*, at low temperatures) and without (*R*3*m*, at high temperatures) octahedral tilts. More precisely, neutron diffraction [20] and high-resolution x-ray diffraction [21] studies confirmed the rhombohedral phase and also revealed the presence of the monoclinic phase (assigned to the *Cm* symmetry in ref. 20), providing further support to the idea that the monoclinic phase is not due to the presence of adaptive phases. At the highest values of *x* an antiferroelectric orthorhombic phase is formed. The local disorder can be divided into Pb displacements (off-site *A* cations) and different positions of the *B* cations, Zr and Ti [22,23,24,25,26]. The role of a disordered *B* cation position was studied in detail in ref. 25 in a tetragonal $Pb(Hf_{0.40}Ti_{0.60})O_3$, a ferroelectric closely related and isostructural to PZT.

The present study focuses on tetragonal PZT single-crystal with *x* = 0.35 and its Brillouin zone-centre modes. Raman scattering is very sensitive in probing deviations from the average symmetry and has been frequently used to address the changes in the vicinity of the MPB [27,28,29] ceramics and defected PZT fibres [30]. In contrast to classical piezoelectric perovskites the structural disorder inherently present in PZT is the key to understanding the phase transformations in this system and its use as a piezoelectric material.

## Experimental

Single crystals of $Pb(Zr_{0.35}Ti_{0.65})O_3$ were grown by a top-seeded solution growth technique using a mixture of PbO and $B_2O_3$ as flux. A crystal platelet of pseudocubic (001) orientation with dimensions 2.2 mm × 1.9 mm × 0.22 mm was prepared with the largest (001) faces mirror-polished. The two long edges are parallel to the pseudocubic *a* and *b* axes and the short edge is parallel to the c axis.

Raman measurements were performed using a Jobin-Yvon T64000 spectrometer consisting of a double monochromator coupled to a third monochromator stage with 1800 grooves per millimeter grating (double substractive mode). The acquisition time was adjusted to have a sufficient signal-to-noise ratio. A liquid nitrogen cooled charge-coupled device detector was used to count photons. All measurements were carried out under a microscope in backscattering configuration. The Raman spectra were excited using a continuous wave solid-state laser (wavelength 532 nm) and a He-Ne laser (wavelength 632.8 nm). The laser beam power on the sample surface was 500 μW, and the diameter of the laser beam spot was approximately 2 μm. The laser light was focussed on a sample surface by a long focal length objective (50×, N/A =0.5). Linkam stages (model TS-1500 and THMS-600) and a temperature control unit (model TMS-94) were used for the room, high and low-temperature measurements. The spectra were not divided by the Bose-Einstein thermal factor as there is no a priori knowledge justifying the assumption that the signal is solely due to first-order Raman scattering. The Raman spectra were analyzed using Jandel Scientific PeakFit 4.0 software.

## Results

Raman-active modes and the atomic displacements spanning their irreducible representations are summarized in Appendix I. The measurement geometries applied in this study are also given in the Appendix I.

### *E*(TO) modes

Figure 1 shows the spectra collected with $x(zy)\bar{x}$ geometry, which reveals only *E*(TO) modes. To address the possible excitation wavelength dependence, measurements were conducted using 532 and 633 nm laser beam wavelengths. The spectra are essentially equivalent. It is seen that the spectra consist of four bands, corresponding to the four *E* modes seen in $PbTiO_3$. The crucial difference is that all the modes, except for the *E*(3TO) mode, are split. The soft *E*(1TO) mode is split into four modes, whereas two are seen in the *E*(2TO) mode. The highest-frequency *E*(4TO) mode is split into four modes. Apart from the splitting of the three *E*(TO) modes there is also a broad background contribution, labelled *E*(BG), seen between 120 and 400 $cm^{-1}$.

Figure 1 shows the Raman shifts as a function of temperature. The splitting is nearly constant as a function of temperature, except in the *E*(2TO) modes, which shows that the two modes merge at around 300 K. Fig. 2 indicates that the peak intensities, including the area of the *E*(BG), increase roughly linearly with increasing temperature. Basically, the intensity of the first-order Raman scattering should increase directly proportionally to $n(\omega)+1$, where $n$ is the Bose-Einstein thermal factor and $\omega$ is the phonon frequency [31]. The other important factor influencing the intensity is the change in structure (bond lengths and angles) as a function of temperature. The *E*(BG) seems to consisted of a nearly continuum series of peaks and thus there is no single peak centre. However, the intensity behaviour is characteristic of the behaviour expected for first-order scattering.

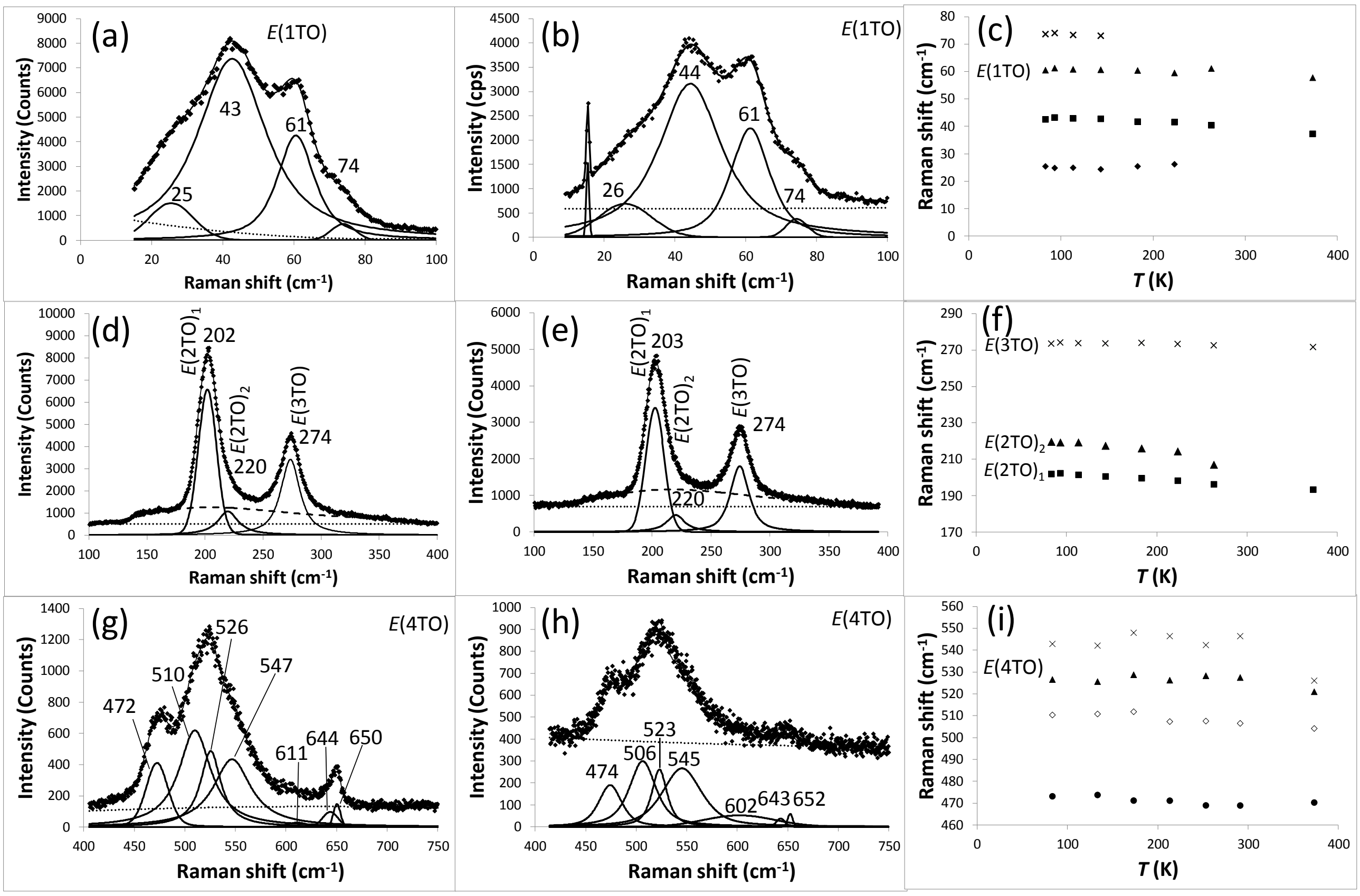

Fig. 1. Raman spectra measured at 83 K temperature with $x(zy)\bar{x}$ geometry which reveals only *E*(TO) modes. Left-hand side panels show data measured using 532 nm excitation wavelength, whereas the middle panels show the data measured using a HeNe-laser (wavelength 632.818 nm). The spectra are very similar, thus showing that the light scattering is truly Raman scattering, not only in the case of the strong peaks, but also in the case of the broad background *E*(BG) shown by the dashed line. Panels (a) and (b) show the splitting of the soft *E*(1TO) mode, panels (d) and (e) show the splitting of the *E*(2TO) mode and the *E*(3TO) mode. Also the highest-frequency peak, labelled *E*(4TO) (panels (g) and (h)) is split. The peaks above 600 $cm^{-1}$ correspond to the $A_1$(3TO) modes and should not be observed in this geometry. Right-hand side panels plot the Raman shifts, (c) shows the soft-mode shifts, panel (f) the *E*(2TO) and *E*(3TO) modes and (i) the highest-frequency *E*(TO) mode as a function of temperature. Except for the *E*(3TO) mode (no splitting) and *E*(2TO) mode (splitting vanishes at around room temperature) no strong temperature dependence of the peak splitting was observed.

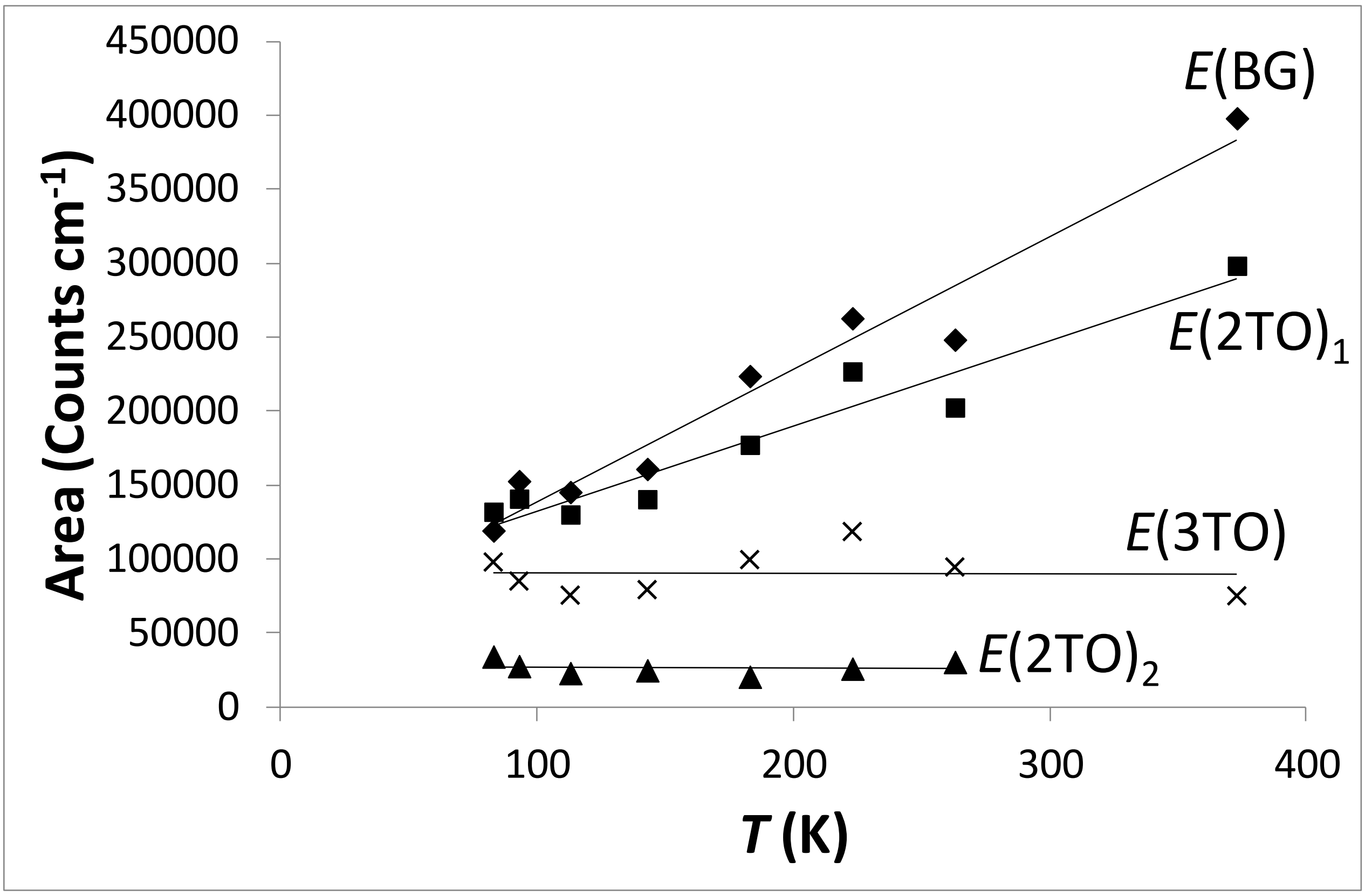

Fig. 2. The area of the broad background feature (labelled as *E*(BG)) seen in Fig. 1 and the area of the *E*(2TO) modes and the *E*(3TO) mode as a function of temperature.

**$B_1$ mode**

In $PbTiO_3$ the $B_1$ mode is degenerate with the *E*(3TO) mode, though this is not required by symmetry. Fig. 3 (a) shows spectra measured at two different geometries, one revealing only the $B_1$ mode, and the second showing only the *E*(TO) symmetry modes. The difference between the *E*(3TO) and the $B_1$ Raman shifts is clear, showing that the degeneracy is obviously broken. Fig. 3 (b) indicates no abrupt changes in Raman shifts, consistent with the notion that there is no low-temperature phase transition. The atomic displacements in the $B_1$-mode are determined by the symmetry alone and involve only the motion of the $O_2$ and $O_3$ oxygens (the basis is $z_{O_2} - z_{O_3}$ and thus this mode breaks the tetragonal fourfold symmetry). The decreasing shift with increasing temperature is due to the decrease and increase of the *c*- and *a*-axes lengths, respectively. No splitting of the $B_1$ mode was evident; in contrast its lineshape was very well fit by a Lorentzian function. This supports the idea that the oxygen network in PZT is very rigid and is well described by the average symmetry. In line with this observation are the neutron diffraction results, according to which the atomic displacement parameters (ADP) of oxygen are normal, whereas the ADPs of cations are anomalous. This is somewhat surprising, as the atomic displacements are parallel to the *c*-axis, suggesting that the anharmonic effects should be strong as oxygen, as expected, has a large vibration amplitude.

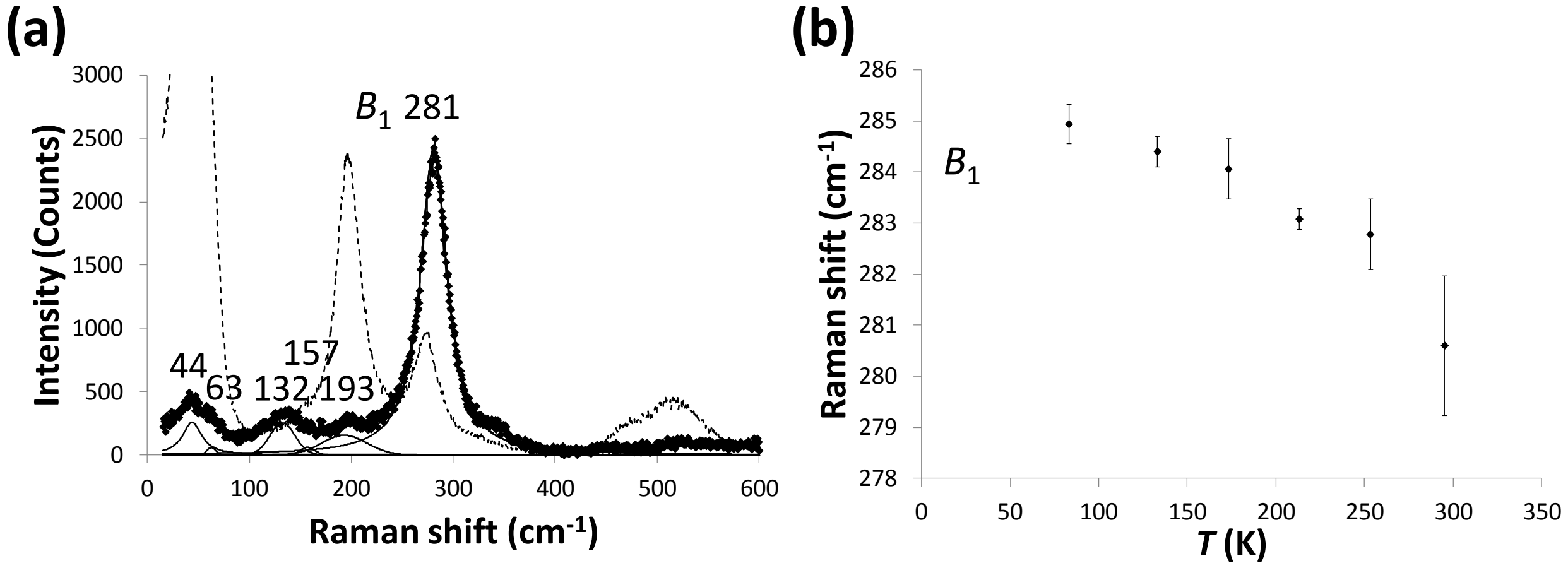

Fig. 3. (a) Raman spectra measured at 295 K temperature with $z[(xy)(\bar{x}y)]\bar{z}$ geometry ideally showing only the $B_1$ symmetry mode (weak features, assigned to the longitudinal $A_1$(LO) and transverse *E*(TO) modes are seen). For comparison, spectrum measured with $z(xy)\bar{z}$ geometry, revealing only *E*(TO) modes, shown by dashed line. In contrast to $PbTiO_3$, the energy difference between the $B_1$ and *E*(3TO) modes is clear. (b) Raman shifts of the $B_1$ mode as a function of temperature. 95% confidence limits are also indicated, reflecting the quality of the fit.

## $A_1$(TO) modes

Figure 4 shows the $A_1$(TO) modes, observed with $x(zz)\bar{x}$ geometry, and plots the Raman shift values. The $A_1$(TO) mode peaks are seen to be asymmetric. In semiconductors, a common cause of asymmetry is Fano resonance (see, e.g., ref. 32). In the present case the lineshape is independent of the excitation wavelength, consistent with the insulating nature of the sample. Furthermore, spectra measured with different wavelengths are very similar (compare Fig. 4 (a) to (b) and (d) to (e)) thus showing that the light scattering is truly Raman scattering. The subpeak structure of each $A_1$(TO) mode can be seen in Fig. 4 (g) and its origin is in the anhamonic potential closely related to the ferroelectricity: because of the ferroelectric polarization the atoms feel different forces when moving parallel and antiparallel to the polarization direction [9,29,33]. A simple way to model the anharmonicity is to convert the problem into a one-dimensional anharmonic oscillator [29]. The $A_1$(1TO) peak became nearly symmetric at room-temperature, indicating that the high-frequency subpeak had merged with the fundamental peak. At 400 K a low-frequency subpeak at 102 cm$^{-1}$ was revealed by the curve fitting.

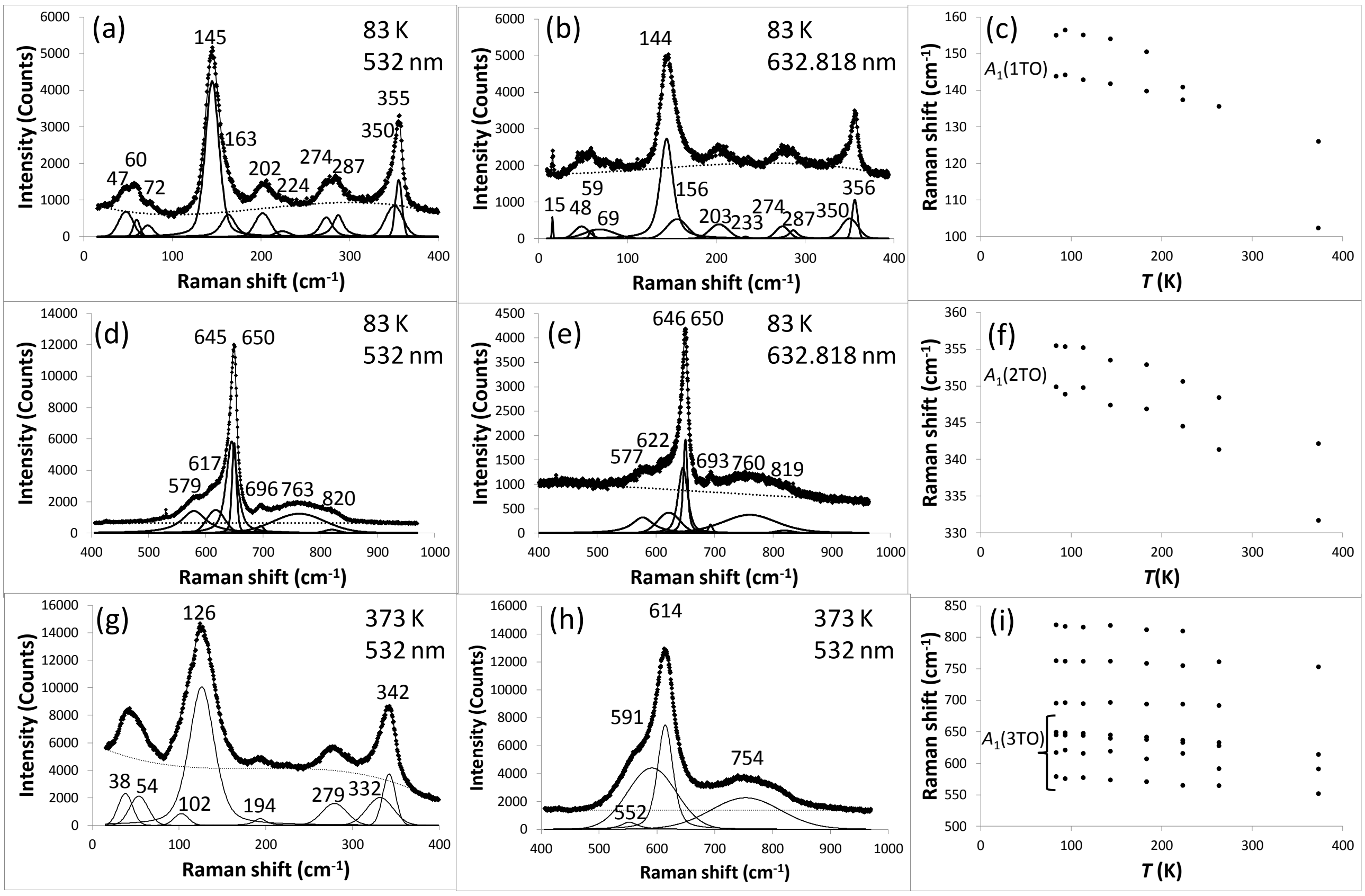

Fig. 4. Raman spectra measured with $x(zz)\bar{x}$ geometry at 83 and 373 K which ideally reveals only $A_1$(TO) modes. Right-hand side panels plot the Raman shifts as a function of temperature. Two excitation wavelengths, 532 and 632.818 nm, were utilized. Panel (a)-(c) show the fundamental peak of the $A_1$(1TO) mode (at around 145 cm$^{-1}$ and the high-frequency mode at 160 cm$^{-1}$, revealed by the asymmetry of the peak) and the $A_1$(2TO) modes (two peaks at around 350 cm$^{-1}$, panels (a), (b), (f) and (g)). Panels (d), (e), (h) and (i) show the splitting of the $A_1$(3TO) mode (peaks below 650 cm$^{-1}$). We note that the lowest-frequency peaks have nearly the same frequency as the strong $E$(1TO) peaks, and also the peaks corresponding to the $E$(2TO), $E$(3TO) and $B_1$ are observed. This is since it is difficult to perfectly align small crystals inside cryostat and since the local structure deviates from the perfect tetragonal structure.

The $A_1$(1TO) mode softens with increasing temperature more than the $E$(1TO) mode. Also the softening of the $A_1$(2TO) and $A_1$(3TO) modes is obvious. The lowest-frequency mode, $A_1$(1TO), contains a strong peak at 145 cm$^{-1}$ (83 K) and a higher frequency subpeak at 163 cm$^{-1}$ (83 K). At elevated temperature lower-energy subpeaks also became more intense. Note that the higher-frequency subpeak disappeared at room temperature. A similar type of subpeak structure is seen in the $A_1$(3TO) mode: the strongest peak in Figs. 4 (d) and (e) could not be properly fitted by a single peak. The change in the subpeak intensities was revealed by peak fitting and is notable in the $A_1$(3TO) mode: the low-frequency shoulder is much more pronounced at 373 K than at 83 K. Features above the fundamental peak at around 650 cm$^{-1}$, as in the $A_1$(1TO) mode, are a sign of structural disorder. In a simple anharmonic crystal with a double-well potential one expects a fundamental peak and a series of lower-energy peaks as the distance between adjacent levels (indexed by $m$) becomes smaller with increasing $m$. Thus, the higher-frequency peaks do not fit with the simple double-well picture but seem to have a different origin. Somewhat peculiarly, the high-frequency shoulder of the $A_1$(1TO) mode seems to disappear at approximately 250 K (Fig. 4 (c)). In ceramics this mode has been reported in several studies [13,34,35]. In a recent first-principles-based simulation the coupling between the $A_1$ and an antiferrodistortive (AFD) mode was considered and found that the higher-frequency $A_1$ peak mainly originates from an interaction between the $A_1$ and an AFD mode in which the atomic displacements are perpendicular to the polarization direction. The simulation indicated a transition from *P*4*mm* to *I*4*cm* (which involves oxygen octahedral tilting) at around 130 K, and that the double peak structure was apparent at a temperature interval between 100 and 500 K but not above or below. Somewhat surprisingly, the aforementioned coupling was also found when a virtual (a kind of average of Ti and Zr) atom was used instead of randomly distributed Ti and Zr atoms. We note that the high-frequency mode is absent in pure $PbTiO_3$, but with just a small *A*-cation substitution [$(Pb_{0.97}Nd_{0.02})TiO_3$ [34,35]], in addition to Zr alloying, made the mode visible. This suggests that defects at the *A* and/or *B* cation sites are required to introduce the mode. Another difference between the presently reported high-frequency mode and the simulated mode is that. instead of having a temperature window at which the mode appears, it is present up to rather high temperatures, and is actually most clearly observed at low-temperatures. This could be due to the stiffer AFD mode (when compared with the MPB PZT), though it requires further studies.

**Structural model**

To explain the observed Raman peak splitting and deviations from the analysis based on an idealized crystal a model in which the short-range order differs from the average long-range order is constructed. We first note that the number of peaks is significantly large: the *E*(1TO) mode is split into four, the *E*(2TO) mode into two, peculiarly the *E*(3TO) mode seems not to be split, and the *E*(4TO) mode is split into four peaks, whereas no splitting is expected by the symmetry analysis, and no peak splitting of the pure oxygen mode ($B_1$ mode) is seen. A second observation is that the separation between the peaks remains approximately constant as a function of temperature. Thus, no indication of a low-temperature phase transition was found. As a third observation we note that the $A_1$(TO) modes are split, and the relative intensities of the peaks strongly change as a function of temperature (e.g., compare Figs.4 (a) and (g)). A fourth peculiarity is the high-frequency band approximately centered at 750 cm$^{-1}$, well above the $A_1$(3TO) peak. We emphasize that the measurement geometry $x(zz)\bar{x}$ does not allow the observation of $A_1$(LO) modes. Nevertheless, the mode frequencies are in the same frequency range as the longitudinal $A_1$(3LO) mode (which is also split, the Raman shifts being 745, 777 and 818 cm$^{-1}$). Thus, though the observation of the LO modes might be due to a domain(s) related to the prevailing crystal by a 90° rotation about the *b* axis we seek another explanation. The puzzle is that no signs of $A_1$(1LO) (164 cm$^{-1}$) or $A_1$(2LO) (448 cm$^{-1}$) modes were seen, which shows that twinning does not provide a satisfactory explanation.

Trivial homogeneous symmetry-lowering, preserving the same number of atoms in the primitive cell, does not provide an explanation. In the PZT solid-solution system the *B*-cation site is typically assumed to be statistically occupied by Zr and Ti ions (i.e., no ordering occurs, consistent with the diffraction experiments). This assumption implies a statistical variation in the spatial composition, which at first sight should correspond to broadening of the Raman peaks, in clear contrast to the present experimental results. It is worth noting that the splittings, including peak positions and relative areas, were independent from the measurement spot and excitation wavelength. Measurement geometries $x(zy)\bar{x}$ and $x(yz)\bar{x}$, obtained by rotating the sample thorugh 90°, gave essentially identical spectra (*E*(TO) modes), as expected for 4*mm* point group symmetry.

In titanium-rich PZT no superlattice reflections are revealed by neutron or x-ray diffraction. Thus, signals from differently ordered local areas average away. An electron diffraction (ED) study [24] revealed large changes in diffuse streaking in PZT as a function of nominal composition. The ED results were interpreted in terms of a model where tetragonal and rhombohedral regions possess short-range ordered areas with correlated lead displacements. The areas grow in size once the MPB region is approached. Diffuse scattering was found to be strongest along high-order zone axes, as was demonstrated by recording a diffraction pattern along the [100] and [111] directions on rhombohedral PZT. This implies that the Pb displacements in a line parallel to any particular ⟨111⟩ direction are correlated, and that there is no correlation between neighbouring lines (see Figs. 5 and 6 in ref. 24). In tetragonal PZT the x-ray Bragg peak width of the 200 reflections was larger than the 002 reflections [35]. In principle, changes in the correlation length of the Pb displacement order correspond to changes in entropy and thus affect the specific heat. Specific heat measurements have been performed as a function of temperature between 300 and 1.8 K for compositions around and at the MPB with the result that the Debye temperature is highest when $x = 0.53$ [36]. High-temperature specific heat measurements, between 313 and 1073 K, were conducted for tetragonal PZT, which revealed a large decrease in the maximum of specific heat with increasing Zr-content [37]. This was interpreted as originating from the higher degree of close-packing of the Zr rich samples. It is hard to extract the Pb displacement contribution to the specific heat data and thus more direct techniques are required. Both ED and x-ray diffraction results indicate that there is a well-developed long-range order along the ferroelectric polarization direction, whereas there are poorly correlated, short-range ordered areas along directions perpendicular to the ferroelectric polarization direction. To understand the deviation from the ideal selection rules (assuming perfect tetragonal symmetry) it is also worth stressing that the atomic displacements of cations deviate significantly from the positions given in Table 1 in Appendix I. In contrast, the oxygen positions seem to obey the tetragonal long-range symmetry, as the nearly ideal behaviour of the $B_1$ mode suggests. The symmetry breaking has two effects: the number of Raman-active modes is increased and the strict division into modes with atomic displacements only along the *a* or *b* or *c* axis is no longer valid. Correspondingly, the selection rules are not strictly obeyed.

The results are in line with the present Raman results, which indicate that there are different locally ordered areas which average out to give the effective tetragonal *P*4*mm* symmetry. The experiments further show that PZT belongs to the realm of statistical mechanics, as textbook models for displacive perovskites do not work. We show that methods analogous to the treatment of magnetic materials (see, e.g., refs. 39-41]) possess a large potential in disordered ferroelectric solid-solutions. We introduce a two-dimensional statistical model capable of explaining the observations and to further explain the composition-dependent transitions to lower-symmetry phases. To keep the model tractable, one structural block is introduced. We consider ⟨110⟩ Pb displacements and statistically distributed Ti and Zr atoms. Appendix II describes the mean-field model applied to generate the Zr/Ti and Pb-displacements shown in Fig. 5. The mean-field describes the Pb-Pb interactions, which results in large, uniform areas (coloured blue in Fig. 5). Without the mean field the Pb-displacements would be nearly random also in titanium rich areas, as one could intuitively expect and as was confirmed by computing simulations with no mean field. According to the simulations carried out for different compositions, the number of different configurations significantly increases in the vicinity of the MPB (see Fig. 2 in Appendix II) so that an external electric field can favour a small number of states over the others. This results in a strongly increased electromechanical response, as is known to occur in this composition range [12]. Note that, according to an experimental study, 90° domains in Ti-rich PZT hardly switch, whereas the domains in the two-phase region do switch [38]. The present model accounts for the development of 90° domains and their easy motion MPB compositions.

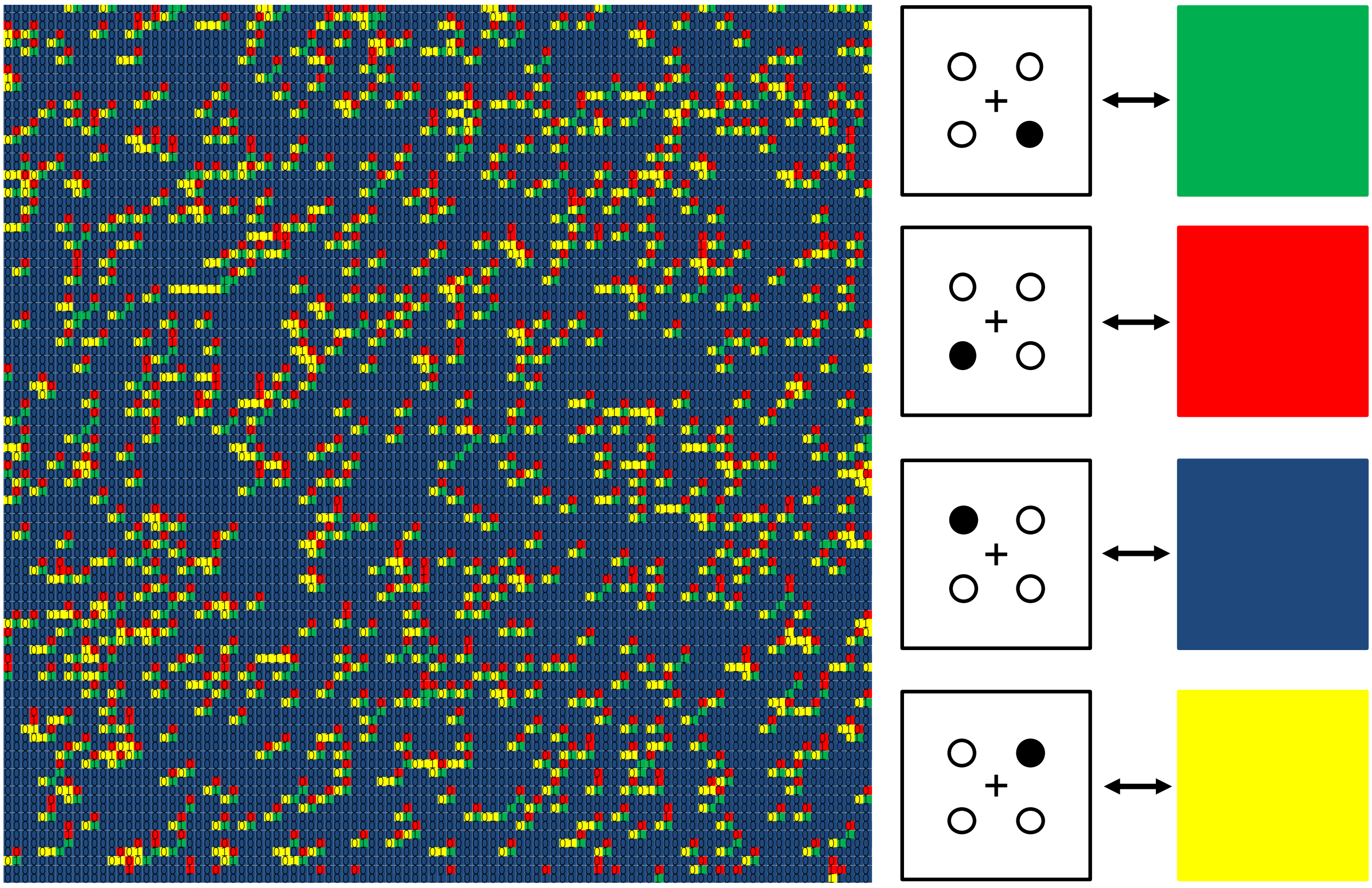

Fig. 5. Distribution of cells for *x*=0.20 and the correspondence between the colours and Pb-displacements. On the local scale, numerous different structures can be identified (compare to Fig. 1(a) in Appendix II). At small Zr concentrations the local symmetry-breaking correlates with Zr sites (denoted by 1), though it is worth noting that also the Ti cells (denoted by 0) have Pb displacements different from the blue matrix.

To fulfil the criteria set by long-range order three other cells are formed by applying 90° rotation. By applying translational symmetry operations an infinite number of structures can be generated from the cell. Different choices correspond to different nearest-neighbour distances between different atomic pairs. Focus is put on the Pb-Zr/Ti distances, see Table 1 in Appendix II. These different short-range orders correspond to different normal modes. Since the distribution of Zr and Ti is not ordered, no long-range order deviating from *P4mm* symmetry is observed: in diffraction experiments peak broadening and anomalous ADPs are signatures of locally ordered areas. A frequently used method to deal with anomalous Pb ADP parameters is to place them in lower-symmetry positions. In refs. 17 and 26, Pb was allowed to move along the [110] directions. The shift from the origin increased with increasing Zr-content and extrapolated to zero for x=0. Examples of different symmetries resulting in different combinations of the cell are shown in Fig. 1 in Appendix II. In the model each of the four Pb-*B* distances can have one of the three available values. If the Pb-*B* distance would be independent of the ion occupying the *B* site, as a first approximation the peak intensities corresponding to each Pb-*B* distance configuration would be proportional to the number of the ways the equivalent configuration occurs statistically (two configurations being equivalent if one is obtained from the other through a rotation or reflection). However, whether the *B* cation is Zr or Ti affects the probability of the occurrence of a given configuration. Raman spectroscopy is very sensitive to local distortions, and the distortions presented in Fig. 1 (Appendix II) break the long-range *P*4*mm* symmetry. Table 1 tabulates the irreducible representations for Raman activity due to the Pb atoms in the case of three dimensional space groups. As is seen, the number of Raman-active modes is significantly increased by the symmetry breaking. As is obvious, the number of totally-symmetric modes is significantly increased, consistent with the experimental observations summarized in Fig. 4 (note especially the high-frequency features, panels (d), (e), (h) and (i)).

Table 1. Correspondence between the plane and space groups and the irreducible representations spanned by Pb displacements. Note the significant increase in the number of Raman-active modes due to the presence of the different cells and that also $B_2$ symmetry modes are activated.

| Plane group | Space group | Total irreducible representation | Raman active representations |
|---|---|---|---|
| *p*4*mm* | *P*4*mm* | $2A_1+A_2+B_1+B_2+3E$ | $A_1+A_2+B_1+B_2+2E$ |
| *p*4*gm* | *P*4*bm* | $2A_1+A_2+B_1+B_2+3E$ | $A_1+A_2+B_1+2E$ |
| *p*2*mm* | *Pmm*2 | $3A_1+3A_2+3B_1+3B_2$ | $2A_1+3A_2+2B_1+2B_2$ |
| *p*1 | *P*1 | 12*A* | 9*A* |
| *pm* | *Pm*11 | 6*A*'+6*A*'' | 4A'+5A'' |
| *cm* | *Cm*11 | 6*A*'+6*A*'' | 4A'+5A'' |
| *c*2*mm* | *Cmm*2 | $4A_1+2A_2+3B_1+3B_2$ | $3A_1+2A_2+2B_1+2B_2$ |

The somewhat surprising result that the *E*(TO) mode peaks are split into well-separated peaks, instead of broad, smeared peaks can now be explained: in the nanometer length scale PZT consists of four types of structural blocks and only three nearest-neighbour Pb-*B* distances are allowed. The model is well consistent with the *E*(1TO) and *E*(4TO) peak splitting, whereas at first sight the *E*(2TO) and *E*(3TO) modes behave slightly differently. Our interpretation is that in this frequency region the splitting contributes mainly to the broad background (BG). The rather abrupt edge at around 135 $cm^{-1}$, seen in Figs. 1 (c) and (d), suggests that the model explains the BG contribution too.

Thus, in our model PZT is consists at this composition of areas possessing different short-range orders. This is consistent with the two-phase model, as applied to MPB PZT (for instance, in refs. 16, 17, 20, 21, and 26) and Zr-rich PZT [16]. It is worth noting that the Bragg reflections of the *Cm* phase are much broader than the reflections corresponding to the *R*3*c* phase [17], in line with the model. The statistical occurrence of each area depends on the composition and applied stimuli, such as stress or electric field.

Recently, the effects of an in-situ sub-coercive electric field neutron and x-ray diffraction studies have been applied to PZT (see, e.g., refs. 42 and 43). We believe that our model provides further insight into the weak-field characteristics of PZT.

## Conclusions

The short and long-range order of a $Pb(Zr_{0.35}Ti_{0.65})O_3$ single crystal were studied by Raman scattering. In strong contrast to $PbTiO_3$, the splitting of transverse *E*-symmetry modes reveals that there are different, locally ordered regions, which on average give tetragonal *P4mm* symmetry. A random distribution of Ti and Zr at the *B*-cation site and Pb ions displaced from their ideal positions are the types of defects making PZT different also from the viewpoint of the phase transition mechanism. In contrast to $PbTiO_3$, a simple expansion of the free energy in terms of order parameter is not sufficient. A statistical model, based on a single structure block, was found to be capable of explaining the observed peak splitting and phase transformations to a large number of different phases in the vicinity of the MPB. Thus, the results support the earlier idea that Pb ions respond readily to an applied electric field, here modelled as a transition between four available positions in each cell. The results further suggest that different crystal symmetries can be stabilized by cooling the crystal through the phase transition temperature under an applied electric field. Analogously to $PbTiO_3$, the coupling between local strain and transverse optical *E*(TO) modes is crucial, as is revealed by a flatter frequency-dependence on temperature compared with the $A_1$(TO) modes. Moreover, each transverse $A_1$(1TO) mode was split by anharmonic interactions as revealed by their temperature-dependent intensities. Thus, the origin of the peak splitting is different in *E*(TO) and $A_1$(TO) modes. No such coupling or anharmonic effects was evident in the $B_1$ symmetry mode in which oxygen ions are displaced along the polarization direction.

## Acknowledgements

Raman measurements were conducted at the Center for Nanophase Materials Sciences, which is sponsored at Oak Ridge National Laboratory by the Scientific User Facilities Division, Office of Basic Energy Sciences, U.S. Department of Energy. The work at Simon Fraser University was supported by the U. S. Office of Naval Research (N00014-11-1-0552 and N00014-12-1-1045) and the Natural Science and Engineering Research Council of Canada. The authors thank Dr. C.M. Rouleau for assistance with the Raman measurements.

**Appendix I: Symmetry considerations**

**Measurement geometries**

Symmetry analysis for a tetragonal lead-zirconate-titanate (PZT) structure, given in Table 1, serves as a starting point for interpreting the Raman spectra.

Table 1. Average crystal structure of the tetragonal lead-zirconate-titanate (space group *P*4*mm*).

| | Wyckoff position | *x y z* |
|---|---|---|
| Pb | 1*a* | $0\ 0\ z_{\mathrm{Pb}}$ |
| Ti/Zr | 1*b* | $0\ 0\ z_{\mathrm{Zr/Ti}}$ |
| $O_1$ | 1*b* | $\frac{1}{2}\frac{1}{2}\ z_{\mathrm{O}_1}$ |
| $O_2$ | 2*c* | $\frac{1}{2}\,0\ z_{\mathrm{O}_2}$ |

The Raman-active modes of the structure given in Table 1 span a representation $3A_1 + B_1 + 4E$. Table 2 gives the atomic displacement spanning the irreducible representations. Totally-symmetric $A_1$ symmetry modes are spanned by the atomic displacements of all atoms parallel to the *c* axis direction, whereas the two-dimensional *E* symmetry modes are spanned by atomic displacements of all atoms perpendicular to the *c* axis (displacement directions can be chosen to be parallel to the *a* and *b* axes). The $B_1$ mode is the only one that is fully determined by symmetry: $O_2$ and $O_3$ atoms vibrate in opposite directions parallel to the *c* axis.

Table 2. Atomic displacements spanning the Brillouin zone-centre $A_1$, $B_1$ and *E* irreducible representations.

| Irreducible representation | Basis |
|---|---|
| $A_1$ | $z_{\mathrm{Pb}}, z_{\mathrm{Zr/Ti}}, z_{\mathrm{O}_1}, z_{\mathrm{O}_2} + z_{\mathrm{O}_3}$ |
| $B_1$ | $z_{\mathrm{O}_2} - z_{\mathrm{O}_3}$ |
| *E* | $x_{\mathrm{Pb}}, x_{\mathrm{Zr/Ti}}, x_{\mathrm{O}_1}, x_{\mathrm{O}_2}, x_{\mathrm{O}_3}$ |
| *E* | $y_{\mathrm{Pb}}, y_{\mathrm{Zr/Ti}}, y_{\mathrm{O}_1}, y_{\mathrm{O}_2}, y_{\mathrm{O}_3}$ |

Table 3 gives the conditions set for the incoming and scattered light polarization required for observation of the first-order Raman scattering.

Table 3. Raman scattering tensors for the $A_1$, $B_1$, $B_2$ and *E* symmetry modes (point group 4*mm*). For the ideal tetragonal structure given in Table 1 no $B_2$ symmetry mode is allowed. $A_1$ and *E* symmetry modes are also infrared-active (phonon polarization is given in brackets).

$$A_1(z)\ \begin{pmatrix} a & 0 & 0 \\ 0 & a & 0 \\ 0 & 0 & b \end{pmatrix} \quad B_1\ \begin{pmatrix} c & 0 & 0 \\ 0 & -c & 0 \\ 0 & 0 & 0 \end{pmatrix} \quad B_2\ \begin{pmatrix} 0 & d & 0 \\ d & 0 & 0 \\ 0 & 0 & 0 \end{pmatrix} \quad E(x)\ \begin{pmatrix} 0 & 0 & e \\ 0 & 0 & 0 \\ e & 0 & 0 \end{pmatrix} \quad E(y)\ \begin{pmatrix} 0 & 0 & 0 \\ 0 & 0 & e \\ 0 & e & 0 \end{pmatrix}$$

In tetragonal PZT, the Raman-active modes originating from the cubic $T_{1u}$ modes are further split into longitudinal and transverse modes, conventionally labelled by assigning (LO) and (TO) after the symmetry label, respectively. This is since, for the ideal structure, in the Brillouin zone-centre modes the atomic displacements are either along the *c* axis direction ($A_1$ and $B_1$ modes) or perpendicular to it (*E* modes, where the atomic displacement in the first mode of the twofold degenerate pair can be chosen to be along the *a* axis direction and the displacement of the other mode can be chosen to be along the *b* axis direction). The modes were further labelled $A_1$(*i*TO), $A_1$(*i*LO), *E*(*i*TO), *E*(*i*LO), and $B_1$+*E*, *i* = 1,2,3. The index *i* increases with increasing frequency [7,9].

In lead titanate, $B_1$ and *E* symmetry modes originating from the cubic $T_{2u}$ modes remain degenerate and there is no LO-TO splitting, though in principle it is allowed for the *E* symmetry mode. Though there are more modes in the present case we still use the same labelling even though it is strictly valid only for the *P*4*mm* symmetry. By combining the restrictions given in Table 3, the momentum conversation rule and the atomic displacements given in Table 2, the geometries revealing the phonon symmetries can be found. Table 4 shows the different set-ups used in this study.

Table 4. Measurement geometries used for measuring the Raman data in this study. All phonons, except for the *E*(LO) modes, can be identified by experiments through backscattering measurements. The *E*(LO) modes can be observed through platelet measurements [9].

| Irreducible representation | Geometry |
|---|---|
| $A_1$(TO) | $x(zz)\bar{x}$ |
| $A_1$(LO) | $z(xx)\bar{z}$ |
| $B_1$ | $z[(xy)(\bar{x}y)]\bar{z}$ |
| *E*(TO) | $x(zy)\bar{x}$ |
| $B_1$ and $A_1$(TO) | $x(yy)\bar{x}$ |

**Monoclinic clusters**

Appendix II gives a statistical model whose essence is that short-range order is taken into account. We give an example of the activation of modes due to the local monoclinic disorder. The Raman scattering tensors for the point group m are given in Table 5. In the case of the *Pm* cluster the matrices refer to the same axes as in the *P4mm* structure, whereas in the *Cm* cluster the relationship between the monoclinic $\boldsymbol{a}_M$, $\boldsymbol{b}_M$, $\boldsymbol{c}_M$ and tetragonal $\boldsymbol{a}_T$, $\boldsymbol{b}_T$, $\boldsymbol{c}_T$ axes is given by $\boldsymbol{a}_M = \boldsymbol{a}_T - \boldsymbol{b}_T$, $\boldsymbol{b}_M = \boldsymbol{a}_T + \boldsymbol{b}_T$, and $\boldsymbol{c}_M = \boldsymbol{c}_T$.

Table 6 tabulates the geometries for which the *A*' and *A*'' modes can be observed in the case of *Pm* and *Cm* clusters. In the case of triclinic clusters all Raman scattering tensor elements are non-zero.

Table 5. Raman scattering tensors for the *A*' and *A*'' symmetry modes (point group *m*).

A’ $\begin{pmatrix} a & 0 & d \\ 0 & b & 0 \\ d & 0 & c \end{pmatrix}$   A’’ $\begin{pmatrix} 0 & e & 0 \\ e & 0 & f \\ 0 & f & 0 \end{pmatrix}$

Table 6. The activity of the *A*' and *A*'' modes of the *Pm* and *Cm* clusters for the measurement geometries (in terms of tetragonal axes) applied in this study.

| Geometry | *A*' representation | | *A*'' representation | |
|---|---|---|---|---|
| | *Cm* | *Pm* | *Cm* | *Pm* |
| $x(zy)\bar{x}$ | Yes | No | Yes | Yes |
| $x(zz)\bar{x}$ | Yes | Yes | No | No |
| $x(yy)\bar{x}$ | Yes | Yes | Yes | No |
| $y(xx)\bar{y}$ | Yes | Yes | Yes | No |
| $y(zx)\bar{y}$ | Yes | Yes | Yes | No |

**Appendix II: Statistical Model**

We describe a two-dimensional statistical model which has some resemblance to the order-disorder model constructed for $KH_2PO_4$ (KDP), in which case one writes a Hamiltonian term *H* to describe the four-proton interaction of the same cluster, $H = \sum_{\mathrm{ijkl}} \sigma_i \sigma_j \sigma_k \sigma_l$ and a mean-field term due to the non-neighbouring protons. Each proton has two positions between the two oxygen atoms connecting two $PO_4$ groups. Because of the protons light mass tunnelling is significant between the two minima thus significantly affecting to the phase transition. Obviously, Pb behaves differently as it forms partially covalent bonds with oxygen. At low temperatures tunnelling between the minima is not significant, though at elevated temperatures higher-energy states became occupied and thus transitions between the minima become more frequent. In contrast to KDP, the approximation of a homogeneous skeleton, formed by *B* cations and oxygen, is not valid for PZT, as the experiments show that the Pb-Zr and Pb-Ti distances differ. As Fig. 1(b) shows, there is one short, one long and two medium long distances so that 6 different Pb-*B* interactions are counted. Fig. 1shows the projection of a distorted unit cell which we call the cell.

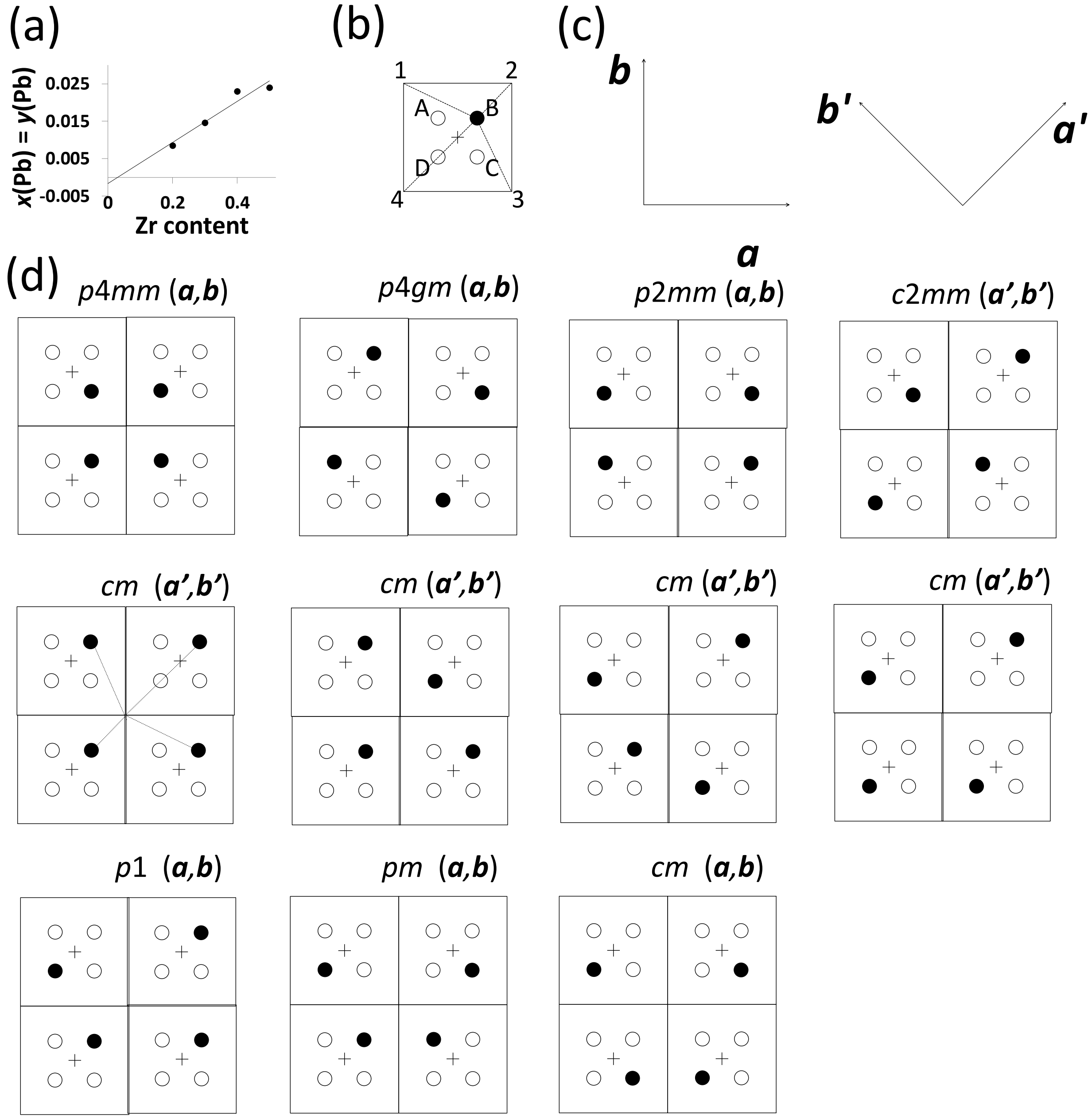

Fig. 1. (a) Pb-displacements (fractional coordinates) along the ⟨110⟩ directions[17}, (b) A cell describing the four Pb (labelled as A, B, C and D) and four *B* cation positions (indexed as 1, 2, 3 and 4) from which an infinite number of short-range order states can be generated by applying fourfold rotation and translational operators. (c) Axes settings specifying the symmetry-element directions characteristic of the planar symmetry groups listed below. (d) Examples of four-cell blocks with planar group symmetries (valid if the blocks are repeated in horizontal and vertical directions). Only projections on the tetragonal *ab*-plane are shown. The cross in the centre of the cell indicates the ideal position of Pb and the four circles, displaced from the cross to <110> directions, are the positions available for Pb. The black sphere indicates an occupied position. For clarity, no other ions are shown. The corners of the cells are the projection points of the *B* cations (Zr or Ti) on the plane perpendicular to the *c* axis. Dotted lines in the diagram on the second row and first column indicate the different Pb-*B* distances. Lattice parameters are relaxed according to the symmetry.

Table 1. Sixteen configurations corresponding to different combinations of Zr (Z) and Ti (T) at the cell corners (Fig. 1) and the corresponding Pb position(s). The mean field removes the degeneracy.

| Configuration | 1 | 2 | 3 | 4 | Pb site |
|---|---|---|---|---|---|
| #1 | T | T | T | T | A, B, C, D |
| #2 | T | T | T | Z | B |
| #3 | T | T | Z | T | A |
| #4 | T | Z | T | T | D |
| #5 | Z | T | T | T | C |
| #6 | T | T | Z | Z | A, B |
| #7 | T | Z | T | Z | A, C |
| #8 | Z | T | T | Z | B, C |
| #9 | Z | Z | T | T | C, D |
| #10 | Z | T | Z | T | B, D |
| #11 | T | Z | Z | T | A, D |
| #12 | T | Z | Z | Z | A |
| #13 | Z | T | Z | Z | B |
| #14 | Z | Z | T | Z | C |
| #15 | Z | Z | Z | T | D |
| #16 | Z | Z | Z | Z | A, B, C, D |

Table 1 shows that several configurations are equally probable. However, after introducing the mean-field contribution due to the Pb-ion displacements the degeneracy is eliminated. The first step in the model construction was the occupation of the *B*-site (indexed by *i,j*) statistically (in the given Zr/Ti concentration) after which the Pb ions were placed according to Table 1 if the configuration corresponding to a lattice site *i,j* was among the configurations #2-#5 or #12-#15, otherwise no Pb was placed. The third step was to compute a mean-field response due to the Pb ions. The fourth step included the treatment of the configurations #1, #6-#11 and #16: in practice the mean field always favoured one Pb site over the others. The distribution of the Pb ions constructed in this way are shown in Fig. 5. An applied electric field would add to the mean field due to Pb ions.

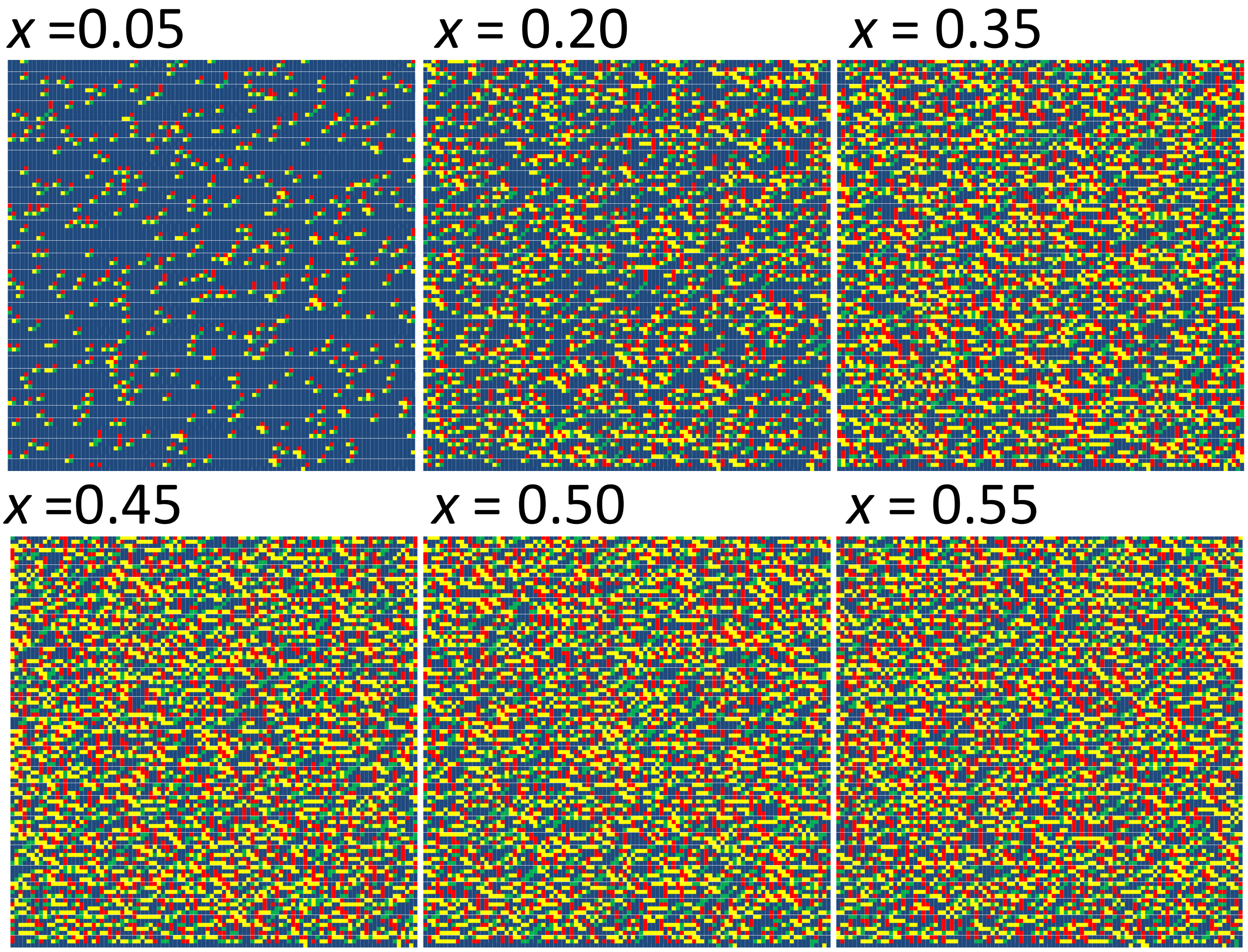

Fig. 2. Distribution of the cells as a function of Zr content *x*. Blue, yellow, green and red squares indicate cells 1, 2, 3 and 4, respectively. Though large uniform areas appear at Ti rich areas, they do not exhibit large deviation from tetragonal symmetry as the Pb-displacements are small when *x* is small (see Fig. 1(a)). In the vicinity of the morphotropic phase boundary a large number of distorted states is available, which explains why the susceptibility of PZT is exceptionally large at this composition region. Each panel has $100\times 100$ cells.

Each of the simulations corresponds to a single domain in the sense that the largest polarization component is perpendicular to the plane of the picture in every cell. An applied field in the vicinity of the MPB favours those local areas in which the Pb-displacements are parallel to the field. Thus, the main polarization component is perpendicular to the plane, whereas there is a small polarization component along one of the $\langle 110\rangle$ directions. In $PbTiO_3$ the $c/a$ axis ratio is large ($\approx$1.064), so that 90° domain reversal involves large structural changes making the domain boundary motion energetically unfavourable. The ratio is close to unity ($\approx$1.01) in the case of MPB PZT, which together with the large number of small polarization components along the $\langle 110\rangle$ directions diminishes the energy barrier for 90° domain reversal. Whereas a large field is required to overcome the internal mean field at small values of $x$, smaller field suffice at $x \approx 0.50$. The four Pb-displacements are equally abundant. Thus at $x \approx 0.50$ an applied electric field can generate domains in which Pb displacements possess long-range order.